\documentclass[twocolumn,showkeys,preprintnumbers,amsmath,amssymb,pra,superscriptaddress]{revtex4}

\usepackage{graphicx}
\usepackage{color}

\usepackage{dcolumn}
\usepackage{bm}

\usepackage{units}

\newcommand{\parenth}[1]{\ensuremath{\left(#1\right)}}

\newcommand{\expec}[1]{\ensuremath{\left\langle #1\right\rangle}}

\newcommand{\jf}[1]{{\color{black} #1}}
\newcommand{\cm}[1]{{\color{black} #1}}
\newcommand{\jdn}[1]{{\color{black} #1}}
\newcommand{\jd}[1]{{\color{black} #1}}

\begin{document}

\title{Markov Chain Monte Carlo Estimation of Quantum States}

\author{James DiGuglielmo}
 \affiliation{%
Institut f\"{u}r Gravitationsphysik, Leibniz Universit\"{a}t Hannover
 and Max-Planck-Institut f\"{u}r Gravitationsphysik (Albert-Einstein-Institute),\\
Callinstrasse 38, 30167 Hannover, Germany
}%
 \email{James.DiGuglielmo@aei.mpg.de}
\author{Chris Messenger}
 \affiliation{%
Institut f\"{u}r Gravitationsphysik, Leibniz Universit\"{a}t Hannover
 and Max-Planck-Institut f\"{u}r Gravitationsphysik (Albert-Einstein-Institute),\\
Callinstrasse 38, 30167 Hannover, Germany
}%
\author{Jarom\'{i}r Fiur\'{a}\v{s}ek}
\affiliation{
Department of Optics, Palacky University, 17. listopadu 50, 77200
Olomouc, Czech Republic
}%
\author{Boris Hage}
 \affiliation{%
Institut f\"{u}r Gravitationsphysik, Leibniz Universit\"{a}t Hannover
 and Max-Planck-Institut f\"{u}r Gravitationsphysik (Albert-Einstein-Institute),\\
Callinstrasse 38, 30167 Hannover, Germany
}%
\author{Aiko Samblowski}
 \affiliation{%
Institut f\"{u}r Gravitationsphysik, Leibniz Universit\"{a}t Hannover
 and Max-Planck-Institut f\"{u}r Gravitationsphysik (Albert-Einstein-Institute),\\
Callinstrasse 38, 30167 Hannover, Germany
}%
\author{Tabea Schmidt}
 \affiliation{%
Institut f\"{u}r Gravitationsphysik, Leibniz Universit\"{a}t Hannover
 and Max-Planck-Institut f\"{u}r Gravitationsphysik (Albert-Einstein-Institute),\\
Callinstrasse 38, 30167 Hannover, Germany
}%
\author{Roman Schnabel}
 \affiliation{%
Institut f\"{u}r Gravitationsphysik, Leibniz Universit\"{a}t Hannover
 and Max-Planck-Institut f\"{u}r Gravitationsphysik (Albert-Einstein-Institute),\\
Callinstrasse 38, 30167 Hannover, Germany
}%

\date{\today}

\begin{abstract}
We apply a Bayesian data analysis scheme known as the Markov Chain Monte Carlo
(MCMC) to the tomographic reconstruction of quantum states. 
\jd{This method yields a vector, known as the Markov chain, which contains the
full statistical information concerning all reconstruction parameters including
their statistical correlations with no a priori assumptions as to the form of
the distribution from which it has been obtained. From this vector can be derived, e.~g.~the 
marginal distributions and uncertainties of all model parameters and also of other quantities
such as  the purity of the reconstructed state. }  We demonstrate the utility of this scheme by 
reconstructing the Wigner function of phase-diffused squeezed states.  These states posses
non-Gaussian statistics and therefore represent a non-trivial case of
tomographic reconstruction.  We compare our results to those obtained through
pure maximum-likelihood and Fisher information approaches.   
\end{abstract}

                             
\maketitle

\section{Introduction}\label{intro}

The tomographic reconstruction \cm{of} quantum states represents
an important laboratory tool in both quantum optics and quantum
information alike. \jf{It} can be applied, for example, to the reconstruction
of the dynamical interaction between quantum systems in a technique
known as \textit{quantum process tomography}.  This latter technique
is important in quantum computing, where the characterization
of quantum gates is essential to the overall quantum circuit
\cite{chuang97,poyatos97,lanyon07}.  Tomography has even been used as
part of the optimization of advanced interferometry as was done in the
preparation of frequency dependent squeezing \cite{chelkowski05}.   

Since the theoretical discovery by Vogel and Risken \cite{vogel89} that the Wigner
function can be reconstructed from homodyne detector data, a number of reconstruction
schemes have been developed ranging from direct inversion of the tomographic
data by means of the filtered-back projection method \cite{smithey93} to
statistical methods such as maximum-likelihood estimation
\cite{hradil97,banaszek99,fiurasek01,lvovsky04,hradil06,rehacek07}.  An important feature  
of maximum-likelihood methods is the guaranteed positive ``semi-definiteness'' of the
reconstructed state.  The result of a maximum-likelihood reconstruction method is either a density
matrix \cite{lvovsky04,hradil06} or a set of parameters \cite{dariano00,ourjoumtsev06}
which have maximized the likelihood functional given a model of the measurement
apparatus and of the parameterized state. \jf{A full analysis of the experimental data, however, 
should also answer important questions regarding  error-bars  on the estimation of the state parameters,
possible correlations amongst the parameters, and error-propagation when using the reconstructed state 
for further calculations of quantities such as the purity of the state or amount of entanglement.
}

\jf{Several methods have been proposed in the literature to put error-bars on the reconstructed
quantum states. For linear reconstruction techniques based on the averaging of sampling functions
over the experimental data \cite{welsch99} one can  calculate the statistical
uncertainties of the estimated quantities by evaluating  variances 
of the linear estimators \cite{leonhardt96}. Another possible approach is to numerically simulate the whole measurement and
reconstruction process many times assuming that the reconstructed state is the true state
that is being measured upon. The error bars are then calculated from the resulting ensemble of 
reconstructed states \cite{dariano99}. Finally, uncertainties on estimates obtained by maximum-likelihood 
method can be determined by evaluating the Fisher information matrix \cite{rehacek08}. 
This latter approach essentially relies on approximation of the likelihood function 
by a Gaussian and becomes  exact only asymptotically in the limit 
of a very large number of experimental data.}

\jf{In the present paper we show that the uncertainties in quantum state estimation can be
consistently determined by using a general and statistically well motivated  
Bayesian analysis scheme known as \cm{the}
Markov Chain Monte Carlo (MCMC)}.  The method is
based on the implementation of a Markov chain to search the parameter space
resulting in a set of \textit{samples from the joint posterior probability density distribution}
on the unknown parameters of the model.
\jf{This technique produces several important results. First, it yields the Markov
chain containing all of the relevant statistical information about the parameter
space.  Second, one can extract a set of marginalized probability
density distributions for each parameter quantifying the degree of uncertainty
on their estimation.  Third, the resulting chain can be used in further
calculations,} \cm{where one can produce probability density distributions on 
quantities such as of the purity or amount of entanglement of the
reconstructed state.}  

This paper is divided into the following sections: in Sec.~\ref{bayes_analysis}
the necessary concepts required from Bayesian data analysis are introduced
and applied to the case of quantum state estimation.  In
Sec.~\ref{qlikelihood_func}, the quantum likelihood function for phase-diffused 
squeezed states is derived.  In Sec.~\ref{mcmc} the Markov chain Monte Carlo 
algorithm is introduced and finally in Sec.~\ref{exp_results} both the technical 
details of the experimental realization as well as the results of the reconstruction are
presented.

\section{Bayesian Data Analysis}\label{bayes_analysis}

Bayes' Theorem prescribes the rule to invert the relationship between the
experimental data already observed and the parameterized model which could have
generated the measured data set.  The theorem reads
\begin{equation}\label{bayes_theorem}
p (\vec{\lambda} \vert D\,,I) = \frac{p(D \vert \vec{\lambda}\,,I)
p(\vec{\lambda} \vert I)}{p(D \vert I)},
\end{equation}
where we use $D$ to represent our data, $I$ to represent our prior information
or model of the experiment and $\vec{\lambda}$ 
to represent a vector of model parameters.  The quantity  
$ p (\vec{\lambda} \vert D\,,I)$ is known as the \textit{posterior
distribution}, $p(D \vert \vec{\lambda}\,,I)$ is known as the \textit{likelihood
function}, $p(\vec{\lambda} \vert I)$ is the \textit{prior distribution}
and $ p(D \vert I)$ is a normalization factor. The \cm{standard application} of Bayesian
analysis is to calculate the posterior distribution given a parameterized
model of the sought after signal, the measured data and prior probability
distributions on the values of the model parameters.  These three elements are
brought together through Eq.~\parenth{\ref{bayes_theorem}}.

 Let us now construct the likelihood function for the quantum estimation problem.
A general measurement on a quantum system  can be described by the
so-called positive operator-valued measure (POVM). Each  possible measurement
outcome $j$ is associated with a POVM element $\Pi_j$ which is a positive
semidefinite operator. The probability of outcome $j$ can be calculated as
$P_j=\mathrm{Tr}[\Pi_j \rho(\vec{\lambda})]$ where $\rho(\vec{\lambda})$ denotes the density matrix of the
measured quantum system that depends on the model parameters $\vec{\lambda}$. 
Since the total probability of some outcome is $1$, the POVM
elements sum up to identity operator, $\sum_j \Pi_j=\openone$.  This generic framework
 in particular encompasses a tomographic reconstruction
of the state $\rho(\vec{\lambda})$  that consists of several different measurements $M$ 
with possible outcomes indexed by $l_M$. Then $j=(M,l_M)$ becomes a
multi-index indicating both the measurement setting and the measurement outcome 
for a given setting.  Let $ n_j $ denote the observed number of measurement outcome $j$ 
and $N=\sum_j n_j$ represents  the total amount of collected data.  
The likelihood function $\mathcal{L}$ is the probability of observation of a particular set
$\lbrace n_j \rbrace$ for a given  $\vec{\lambda}$. It follows that $\mathcal{L}$
is given by a  multinomial distribution and reads
\begin{equation}\label{qlikelihood}
\mathcal{L} = N !\prod_j^N \frac{P_j^{n_j}}{n_j!}.
\end{equation}
In terms of the constituents
of Bayes's theorem the theoretical probabilities  $\lbrace P_j \rbrace$ are
functions of the parameters which are to be determined.  The measured numbers of 
counts  $\lbrace n_j \rbrace$ correspond to the data.

\section{The Likelihood Function for Phase-Diffused Squeezed States}\label{qlikelihood_func}

The phase-diffused squeezed states were first introduced within the context of
continuous-variable
squeezing purification \cm{by} \cite{fiurasek06,franzen06,hage07}. 
They arise  when squeezed states are transmitted over de-phasing  
quantum channels such as optical fibers affected by thermal fluctuations. 
 These states are characterized
by non-Gaussian statistics and therefore represent a non-trivial case for
quantum state tomography.

The Wigner function for phase-diffused squeezed state is given by
\begin{equation}\label{sqz_phdiffused}
W\parenth{x,p} = \frac{1}{2\pi \sqrt{V_x V_p}} \int_{-\infty}^{\infty} \!\!
\exp{\left[ -\left(\frac{x^2_{\phi}}{2 V_x}+\frac{p^2_{\phi}}{2V_p} \right)\right]}\Phi\parenth{\phi}d\phi,
\end{equation}  
where $x_{\phi}=x\cos \phi +p\sin \phi$, $p_{\phi}=p\cos \phi-x\sin\phi$
with $x$ and $p$ as the standard position and phase quadratures
and $\phi$ representing the random phase shifts distributed according
to some probability distribution $\Phi\parenth{\phi}$.  We use $V_x$
to represent the variance of the squeezed quadrature and $V_p$
for the variance of the anti-squeezed quadrature. We normalize the variances such that 
for vacuum state we have $V_x=V_p=1$ and the state is squeezed in $x$ quadrature if $V_x<1$. 
In the experiment we measure several different rotated quadratures $x_\theta$, where $\theta$ 
defines a specific measurement setting.
The theoretical homodyne probability density distribution $p(x_\theta)$ can be
calculated from Wigner function as a marginal distribution. Integration
of $W(x,p)$ over the conjugate quadrature $p_\theta$ yields,
after some algebra
\begin{equation}\label{homodyne_probdist}
p\parenth{x_\theta} = \frac{1}{\sqrt{2 \pi}} 
\int_{-\infty}^{\infty} \!\!  
\frac{1}{\sqrt{\tilde{V}(\phi)}}\exp{\left[ -\frac{x_{\theta}^2}{2 \tilde{V}(\phi)} \right]}\Phi
\parenth{\phi-\theta} d\phi, 
\end{equation}  
where $\tilde{V}(\phi) = V_x \cos^2 \phi + V_p \sin^2 \phi$. 

\jf{The data from each measurement is binned into $L$ bins whose lower
boundaries are defined by $Q_{\theta,l}$. The outer 
bins extend to infinity  and we set $Q_{\theta,1}=-\infty$ and $Q_{\theta,L+1}=\infty$.
  The corresponding theoretical probability
$P_{\theta,l}$ is given by integration of the probability density 
(\ref{homodyne_probdist}) over the bin,
\begin{equation}
P_{\theta,l}(\vec{\lambda})= \int_{Q_{\theta,l}}^{Q_{\theta,l+1}}
p(x_\theta) d x_\theta.
\end{equation}
\jdn{In our experiment, the results from two quadrature measurements were formed into histograms each 
containing a total of $L = 70$ bins. From the perspective of direct data inversion this corresponds to an 
overdetermined system, because we need to estimate only three real parameters, c.~f.~below.}

The POVM elements describing such binned homodyne detection can be expressed as
\begin{equation}
\Pi_{\theta,l}=\int_{Q_{\theta,l}}^{Q_{\theta,l+1}} |x_{\theta}\rangle\langle x_{\theta}| d x_{\theta},
\label{POVM}
\end{equation}
where $|x_{\theta}\rangle$ is an eigenstate of quadrature operator $x_\theta$.
Note that, by definition, the sum of theoretical probabilities over all bins is equal to one,
\begin{equation}
\sum_{l}P_{\theta,l}(\vec{\lambda})=1.
\end{equation}
This is a mathematical expression of the fact that the homodyne detection always yields some
outcome and, after each measurement, one of $n_{\theta,l}$ is increased by one. Put in a
different way, the homodyne  detection is described by a complete POVM  (\ref{POVM}) 
whose elements $\Pi_{\theta,l}$ satisfy the condition $\sum_l \Pi_{\theta,l}=\openone$. }

Assuming the phase noise distribution, $\Phi\parenth{\phi}$, is a zero mean
Gaussian, the state can be completely characterized by just three
parameters $\vec{\lambda} = \{V_x\,,V_p\,,V_{\phi}\}$ where $V_{\phi}$ is
the variance of the random phase shifts.  The quantum log-likelihood function
for the phase diffused squeezed states is finally obtained by taking the
natural logarithm of Eq.~\parenth{\ref{qlikelihood}} giving us
\begin{equation}\label{loglikfunc}
  \Lambda = \sum_{\theta,l} n_{\theta,l}\ln\left[P_{\theta,l}(\vec{\lambda})\right],
\end{equation}  
where, for simplicity, we ignore all terms that do not depend on the parameter values.

\section{Markov Chain Monte Carlo Algorithm}\label{mcmc}

The goal of our Bayesian reconstruction scheme is the calculation of
marginalized posterior distributions on our model parameters $\jd{\vec{\lambda}
= \lbrace V_x,V_p,V_\phi \rbrace}$.  
To this end, the MCMC method can be used to generate samples drawn from the
posterior distribution $p(\vec{\lambda} | D, I)$.  Since this distribution is unknown
one cannot directly sample from it and instead we sample from the distribution that 
is the product of the likelihood and the prior.  The prior is chosen by
considering the possible values of the parameters to be determined.  Since the
parameters to be determined in this case are variances, their values must be
greater than zero.  In order to assume relative ignorance in the value the
parameters could take, we use a prior which only requires the variances to be
positive and satisfy the Heisenberg uncertainty relation, $V_xV_p \geq 1$. 
 Since the likelihood in this analysis is a sharply peaked function 
the choice of uniform priors has negligible effect on the numerical results of the
MCMC~\footnote{Since the parameters are variances the actual priors
should be determined by using Jeffrey's Principle of Invariance.  However, as stated,
the implementation of flat priors on $\vec{\lambda}$ in this case has negligible effect.}.
In this case Bayes's theorem, Eq.~\parenth{\ref{bayes_theorem}},
tells us that the likelihood function is proportional to the posterior distribution and since this 
is a function that we can compute for a given parameter space location we can use 
a standard sampling algorithm such as the Metropolis-Hastings sampler to draw from it.
We describe our implementation of this sampler in Appendix~\ref{mhsampler}.

\section{Experimental Implementation}\label{exp_results}

\begin{figure}[!t!]
\begin{center}
\includegraphics[width = \columnwidth]{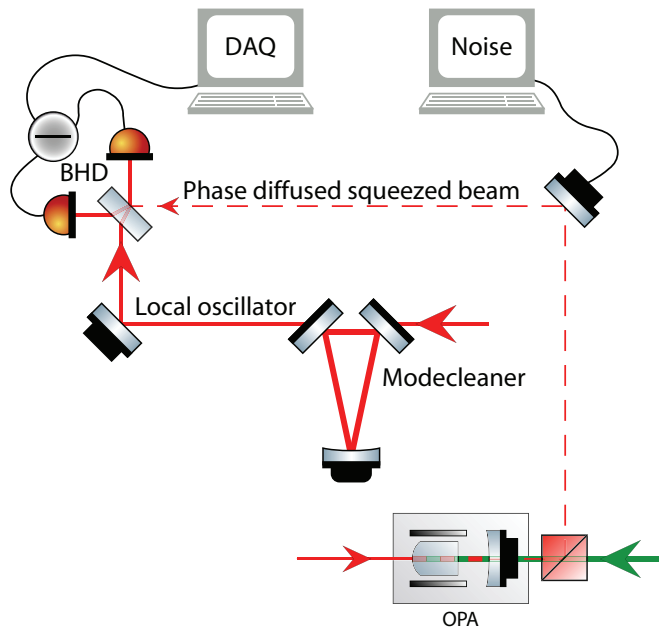}
\caption{(Color online) Experimental setup:  The experimental setup consists of
an optical parametric amplifier (OPA) for the generation of the squeezed vacuum
states, a phase shifter to induce the random phase noise and a homodyne detector
to measure the prepared state.  The mode cleaner was used to increase the fringe
contrast at the balanced homodyne detector.}
\label{exp_setup}
\end{center}
\end{figure}

\subsection{Description of the Experiment}

\begin{figure*}[!t!]
\includegraphics[width = 5.4cm]{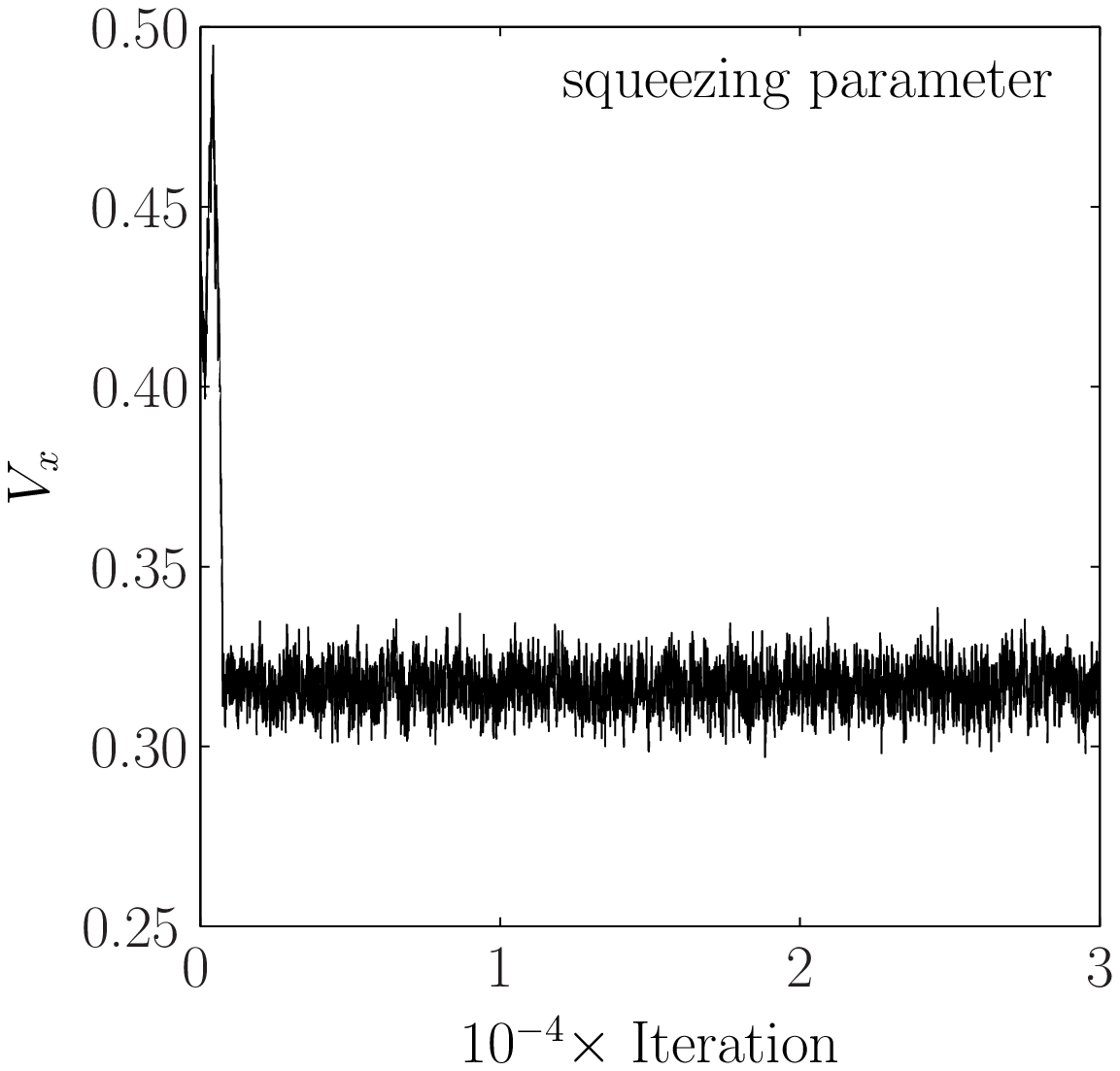}
\includegraphics[width = 5.4cm]{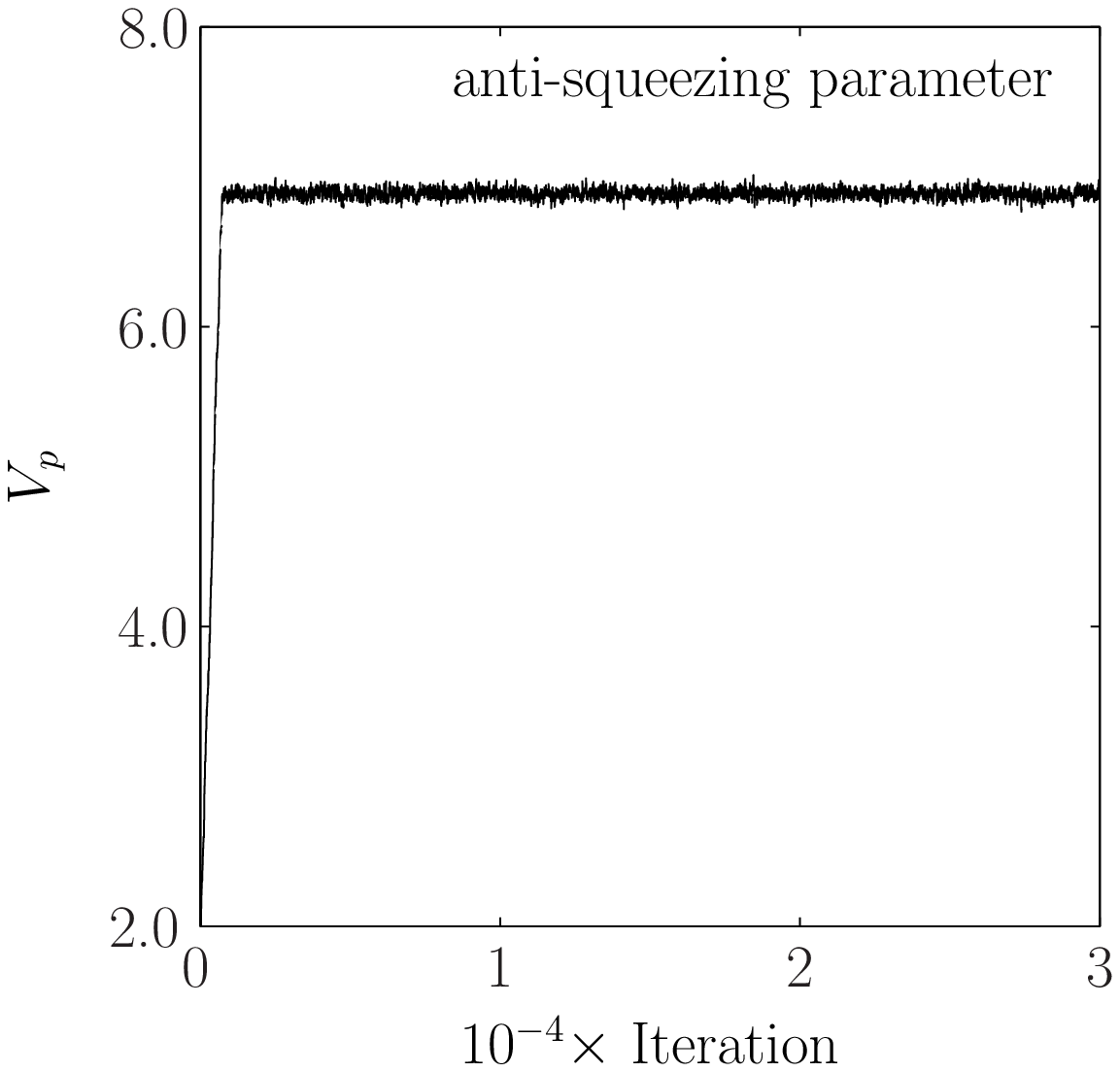}
\includegraphics[width = 5.4cm]{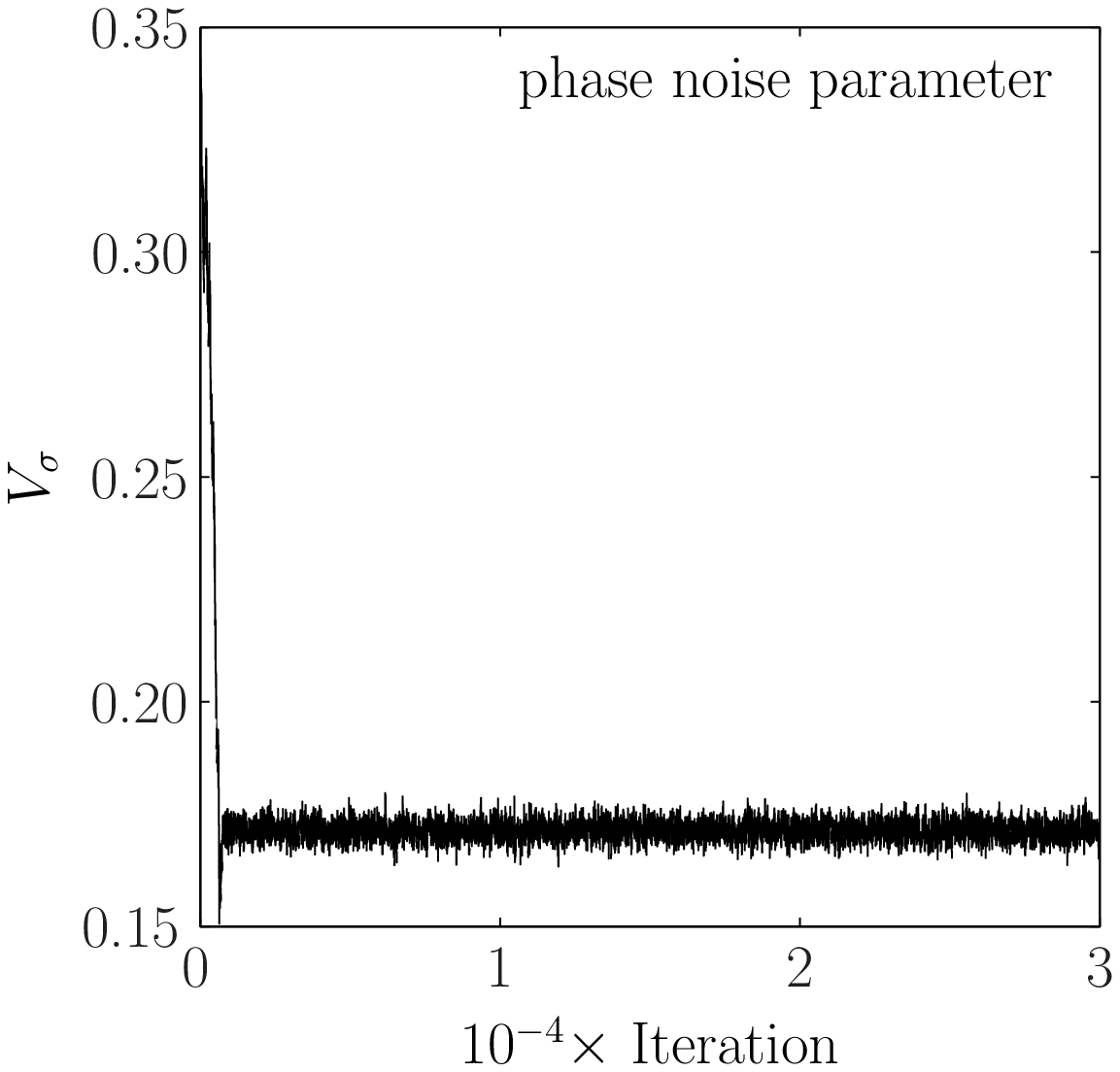}
\caption{Markov chains: This figure depicts the evolution of the
Markov chains where the abscissa represents the iteration number and the
ordinate represents the value of the chain.  After an initial ``burn-in''
period, in which the chains head to their steady-state positions, the
chains eventually converge to a region of parameter space and begin to
\cm{sample from the posterior distribution}.  The proposal distributions were
taken to be Gaussian with the standard deviations $\Sigma_{x}=0.0042$, $\Sigma_{p} = 0.022$,
$\Sigma_{\phi} = 0.0037$ corresponding to the squeezing parameter, the anti-squeezing parameter and the
phase noise parameter, respectively.  The starting values were randomly chosen
with the only constraint that they be non-negative \jf{and obey $V_xV_p \geq 1$}.  The chains settled to their
equilibrium positions with means of $\mu_x = 0.316$, $\mu_p = 6.888$, and
$\mu_{\phi} = 0.171$ corresponding to the squeezing, anti-squeezing and phase
noise parameter respectively.}
\label{mc_chains} \end{figure*} %

\begin{figure*}[!t!]
\includegraphics[width = 5.4cm]{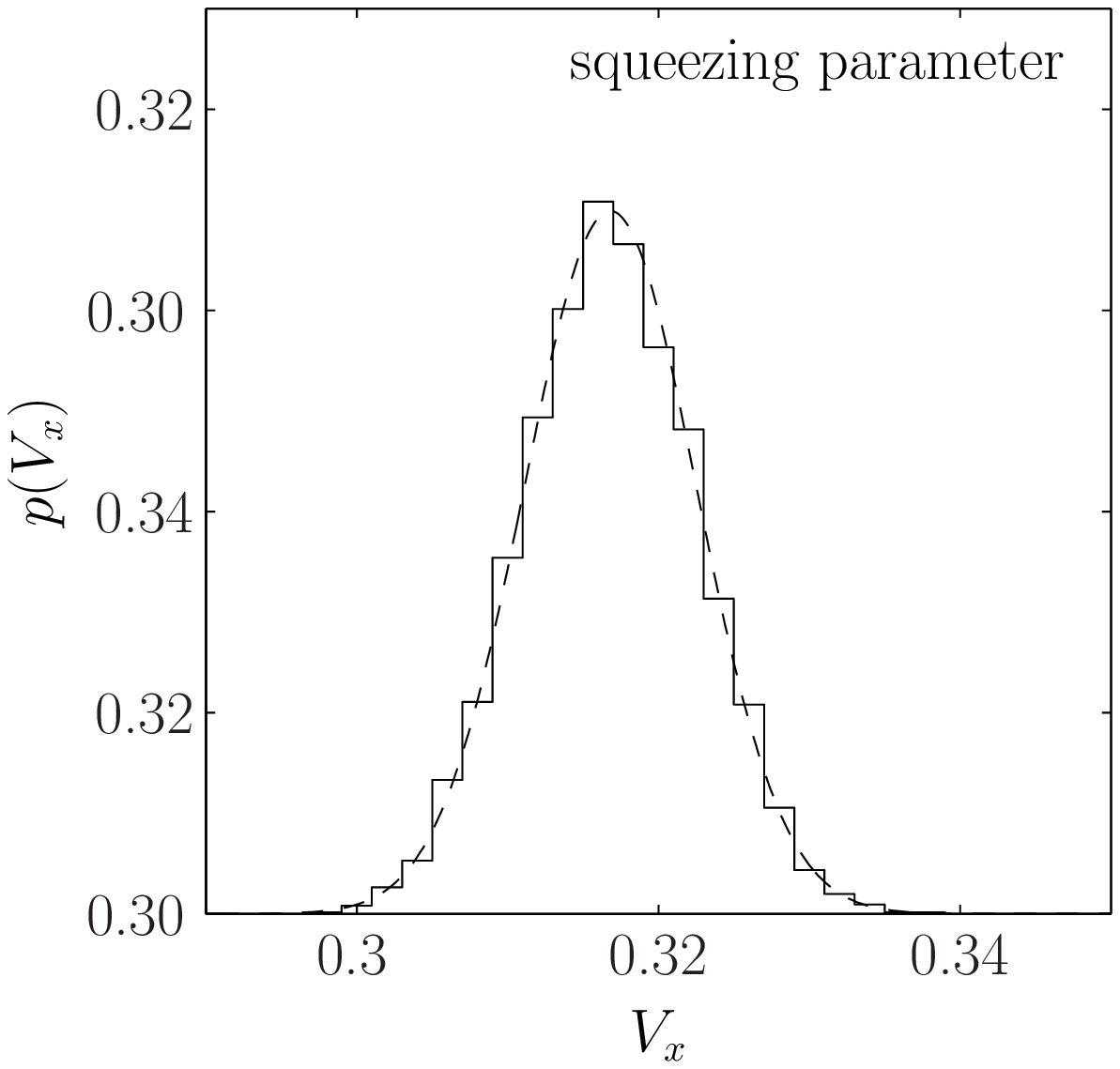}
\includegraphics[width = 5.4cm]{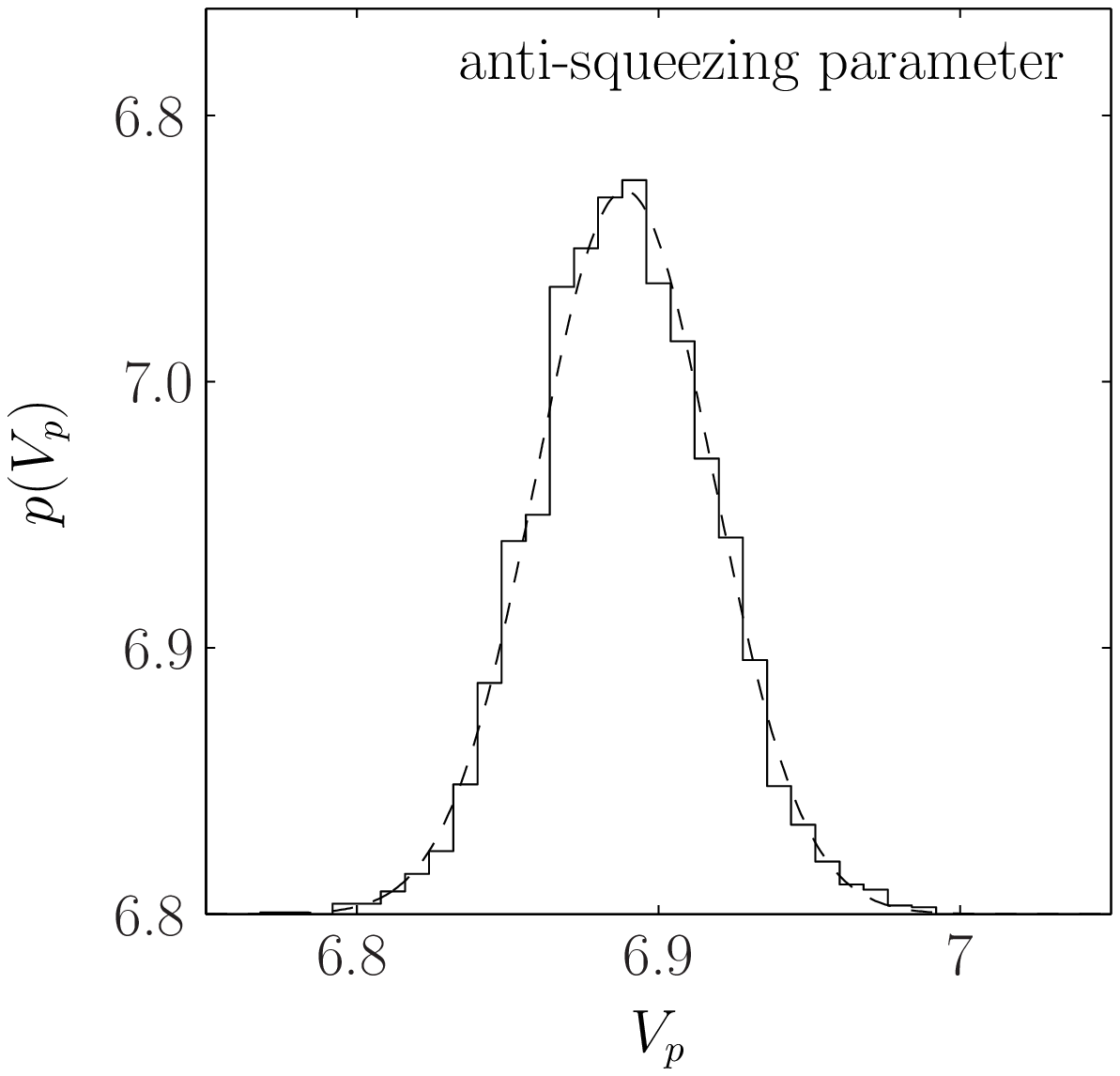}
\includegraphics[width = 5.4cm]{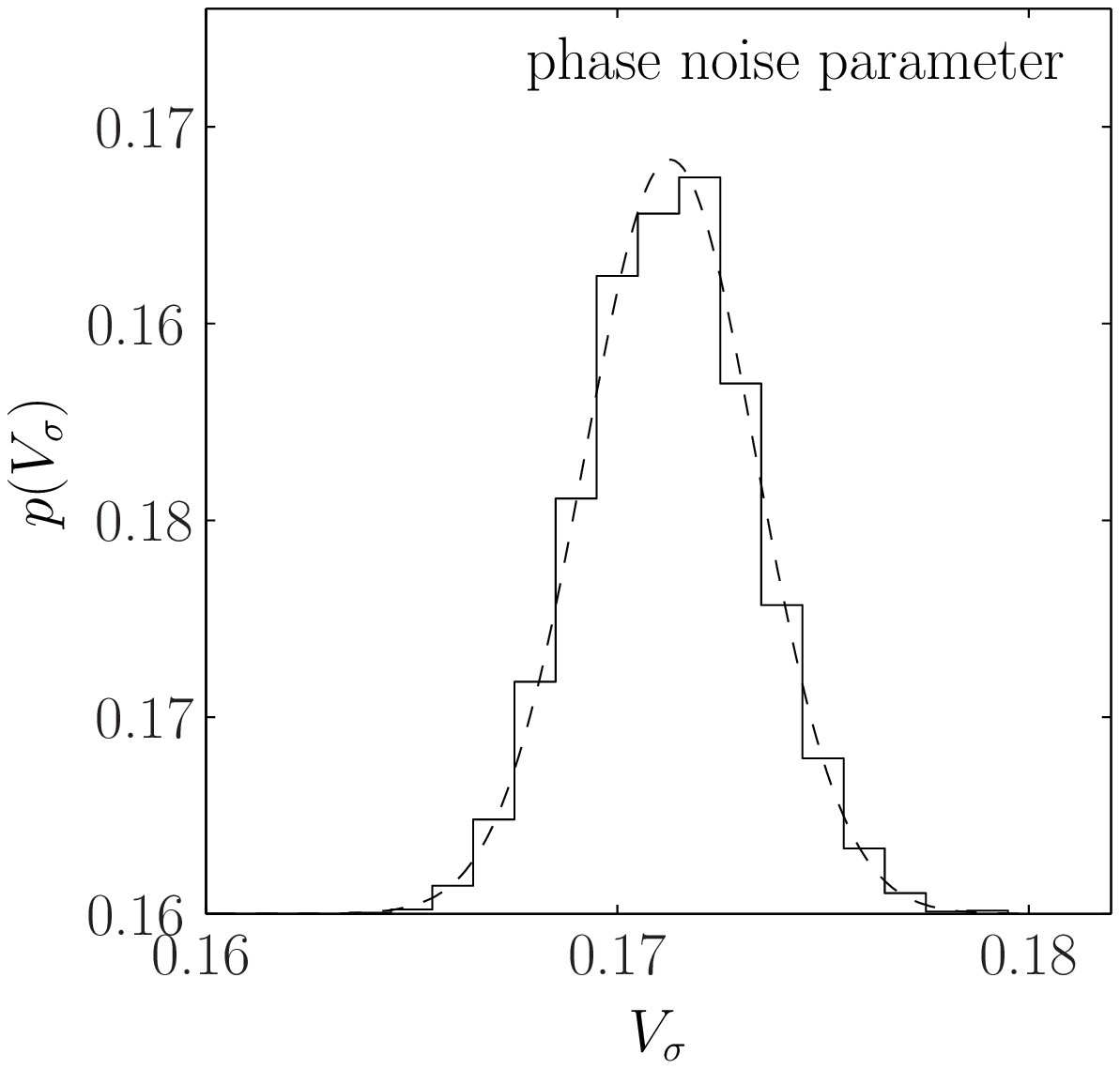}
\caption{Marginalized posterior distributions: Each probability
density was calculated using every tenth point from the marginalized chains
corresponding to 38000 data points.  The standard
deviation of each density are $\sigma_{x} = 0.0056$, $\sigma_{p} = 0.0289$, and
$\sigma_{\phi} = 0.0020$ corresponding to the squeezing parameter, the anti-squeezing parameter and the
phase noise parameter, respectively.  The dotted line represents the result of
the Fisher analysis.}
\label{posterior_dist}
\end{figure*}

Figure~\ref{exp_setup} shows the experimental setup that was used to
prepare the phase diffused squeezed states.  The full details of the setup are
provided in \cite{franzen06} and will be summarized here. The squeezing source
was an optical parametric amplifier (OPA) constructed from a type I non-critically
phase-matched $\textrm{MgO:LiNbO}_3$ crystal inside a standing wave 
resonator, similar to the design that previously has been used in
\cite{chelkowski07}.  The OPA was pumped with \unit[50]{mW} of green light at
\unit[532]{nm} resulting in a classical gain of about 11.  The length of the OPA cavity as well as the phase of
the second harmonic pump beam were controlled using radio-frequency modulation/demodulation techniques.
 The mode cleaner was operated in high finesse mode $\mathcal{F} = 10500$ resulting in a line width of \unit[55]{kHz}.
A non-classical noise power reduction of slightly more than \unit[5.0]{dB} was directly observed with a homodyne
detector in combination with a spectrum analyzer at a Fourier sideband frequency
of \unit[6.4]{MHz}. 

The phase noise was induced by reflecting the squeezed field from a
piezo-electric transducer (PZT) mounted high-reflection mirror that was
quasi-randomly moved.  The voltages applied to the PZTs were produced
as follows.  An independent random number generator produced data
strings with a Gaussian distribution. The strings were digitally filtered
to limit the frequency band to 2--2.5\,kHz.  The output interface was
a common PC sound card with SNR of -110\,dB.  The sound volume was set
to meet the desired standard deviation of channel phase noise.

Homodyne detection confirmed that the squeezing degraded in
the same way when phase noise was increased.  The detector difference currents were electronically mixed with
a \unit[6.4]{MHz} local oscillator. The demodulated signals were then filtered
with a \unit[400]{kHz} bandwidth low-pass filter and sampled with one million
samples per second and 14 bit resolution using a National Instruments analog-digital sampling card.

\subsection{Results of the MCMC}

Figure~\ref{mc_chains} depicts the resulting chains after 40,000 iterations of
the \jf{MCMC} \cm{algorithm}. The abscissa represents the number of iterations of the MCMC whereas the ordinate represents
the parameter values.  After an initial ``burn-in'' period of
approximately 1000 iterations, in which the chain heads towards \cm{equilibrium}, 
the chain converges and begins to sample from the posterior distribution
(which in this Gaussian case also includes the region of maximum-likelihood).
\jd{The development of criteria for the determination of chain convergence is a general problem which has been the
subject of much research \cite{BrooksGelman1998,Eladlouni2006}. The general idea
is to run multiple chains per estimation parameter and monitor their evolution
both within each change and across each change.  Convergence is inferred if all
chains behave consistently. With respect to the case at hand, convergence of the chain can be
inferred by comparing the locations to which the marginalized parameter chains
have settled with the independent measurement of the $V_x$, and $V_p$ parameter
values performed with a spectrum analyzer. In the
absence of such an independent measurement, the criteria in
\cite{BrooksGelman1998, Eladlouni2006} can be used
to infer convergence.}
  
The width of the marginalized chains, i.~e., their standard deviations, 
quantify the degree of uncertainty on the value of each parameter.  By forming
histograms of the chain as a function of each of the parameters we obtain their
marginalized posterior probability distributions as shown in
Fig.~\ref{posterior_dist}.  From these 
posteriors we obtain the following uncertainties on the model parameters: 
$\sigma_x = 0.0056$ for the squeezing parameter, $\sigma_p = 0.0289$ for the
anti-squeezing parameter and finally $\sigma_{\phi} = 0.0020$ for the phase
noise parameter.  The proposal distributions, from which the posterior
distribution samples have been drawn, were chosen to be Gaussians with the
following standard deviations $\jf{\Sigma_x} = 0.0042$, $\jf{\Sigma_p} = 0.022$,
$\jf{\Sigma_{\phi}} = 0.0037$ corresponding to the squeezing parameter, anti-squeezing 
parameter and phase noise parameter, respectively.  These values were obtained 
through manual tuning of the MCMC algorithm. \jd{This is done by adjusting the
individual standard deviations, i.~e.~ $\Sigma_x$, $\Sigma_p$, $\Sigma_{\phi}$,
until the proportion of accepted jumps reaches approximately $44\%$
\cite{gelman04}}.

As an independent test of the posterior standard deviations, we also calculated
the Fisher information matrix \cite{Fisher1925, vallisneri08} given by
\begin{equation}\label{fishermat}
\mathcal{F}_{ij} = \expec{\frac{\partial \Lambda}{\partial
\lambda^{i}}\frac{\partial \Lambda}{\partial \lambda^{j}}}.
\end{equation} 
The inverse Fisher matrix represents the covariance of the posterior probability
distribution for the true parameters $\vec{\lambda}$ as inferred from a single
experiment assuming Gaussian noise and constant priors over the parameter range
of interest.  The calculated standard deviations from the
Fisher matrix are presented in Fig.~\ref{posterior_dist} as well as in
Table \ref{resultstable} where very good agreement is readily seen.  It should be stressed that the
posterior distribution standard deviations obtained from the Fisher matrix
are only valid when assuming Gaussian noise whereas the
posterior distribution standard deviations obtained from the Markov chain is
valid regardless of the form of the posterior distribution.  The results
of the MCMC as well as of the Fisher analysis are compared in Table
\ref{resultstable}.

Figure \ref{loglik_casest} depicts the evolution of the log-likelihood \cm{function 
(Eq.~(\ref{loglikfunc}))} for each iteration of the MCMC. It is seen that as the chains 
evolve \cm{through the ``burn-in'' stage the log-likelihood quickly increases.  After 
approximately $1000$ iterations it has reached equilibrium whereupon parameter space jumps
to higher likelihood values are balanced by jumps to lower likelihood values.}

\begin{figure}
\begin{center}
\includegraphics[width=\columnwidth]{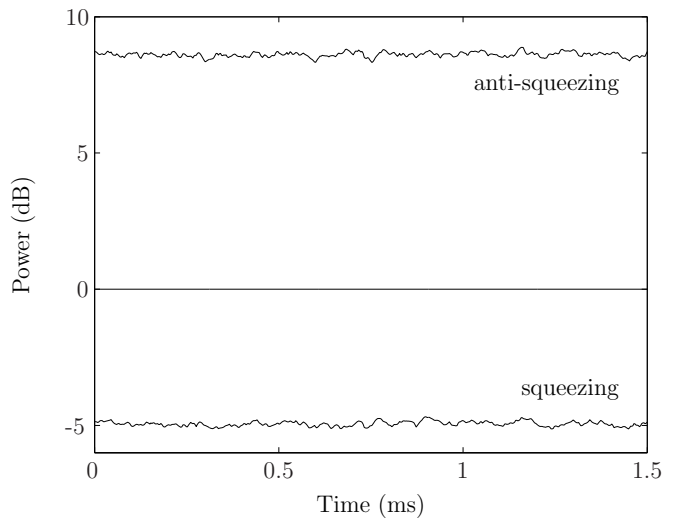}
\caption{Zero span measurement of squeezing and anti-squeezing: The recorded amount of
squeezing and anti-squeezing as measured by a balanced homodyne detector and
spectrum analyzer at a Fourier frequency of \unit[6.5]{MHz}.  
It is seen that approximately \unit[-4.98]{dB}
of squeezing and \unit[8.39]{dB} of anti-squeezing were directly measured without
dark noise correction.  These correspond to variances of $V_{x} = 0.31$ and
$V_{p} = 6.91$, respectively.  The units of the vacuum have been set to 1
corresponding to \unit[0]{dB}.}
\label{specki_measurement}
\end{center}
\end{figure}

\jd{Figure \ref{specki_measurement} represents the spectrum of the squeezed state
before the phase diffusion.  The measured state is seen to have a squeezing
strength of \unit[-4.98]{dB} or a variance of $V_x = 0.31$ and an anti-squeezing
strength of \unit[8.39]{dB} or a variance of $V_p = 6.91$.  These values have
not been corrected for dark noise and were measured using a video bandwidth
(VBW) of \unit[10.0]{Hz}, a resolution bandwidth (RBW) of \unit[100]{kHz} and a
sweep time (SWT) of \unit[1.5]{s}.  These values lie
within the width of the respective posterior distributions obtained from the MCMC analysis.  Furthermore, only
\textit{two} quadrature measurements, each containing just $100,000$ data
samples, were required to obtain these results.  This represents a significant
savings in terms of experimental effort to reconstruct a non-Gaussian state.

\begin{table}
\caption{\label{resultstable} Standard deviations of posterior distributions:
This table compares the standard deviations of the parameter posterior
distributions obtained from the Markov chain Monte Carlo method and from the
Fisher information matrix.  The standard deviations represent the error on the
estimation of the parameter values.  The Fisher matrix returns the actual
standard deviations only in the case of Gaussian noise.}  
\begin{ruledtabular}
\begin{tabular}{ccccc}
 &\multicolumn{2}{c}{Parameter Estimates}&\multicolumn{2}{c}{Uncertainties}\\
 Parameter & MCMC & Maximum Likelihood & MCMC &   Fisher\\ 
\hline
$V_x$    & 0.316 & 0.317 & 0.0056 &   0.0055 \\
$V_p$    & 6.889 & 6.880 & 0.0289   &  0.0294 \\
$V_{\phi}$ & 0.171 & 0.171 & 0.0020  &  0.0020\\
\end{tabular}
\end{ruledtabular}
\end{table}

\begin{figure}[!b!]
\begin{center}
\includegraphics[width = 0.95\columnwidth]{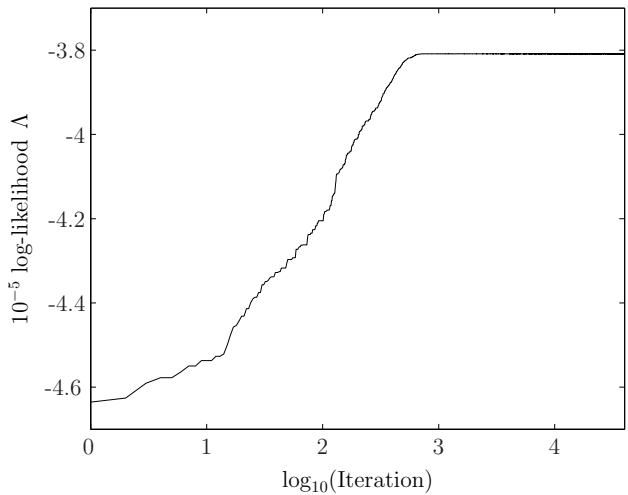}
\caption{Log-Likelihood:  Evolution of the log-likelihood as a
function of iteration. The abscissa represents the iteration number and the
ordinate the log-likelihood value. After approximately 1000 iterations the log-likelihood 
appears to have attained it's maximum value at which point the
chains have reached equilibrium and are sampling from the posterior distribution.}
\label{loglik_casest}
\end{center}
\end{figure}

We also compare our results with that of a pure maximum-likelihood
approach. This was achieved by altering step 3(d) of the
Metroplis-Hastings sampler (see Appendix~\ref{mhsampler}) such that the chain remains at its current
position whenever the likelihood ratio $r$ is less than one.  This forces the chain
to only move to regions of higher likelihood and hence quickly converge on the true maximum-likelihood location.  
The parameter values obtained at the maximum-likelihood were: $\hat{V}_x = 0.317$ for the squeezing parameter,
$\hat{V}_p = 6.880$ for the anti-squeezing parameter and $\hat{V}_{\phi}
= 0.171$ for the phase noise parameter.  These are, however, the only
results obtained from the maximum-likelihood estimation alone.  Since the MCMC
chain contains a complete statistical description of the parameters, the statistical error on
the reconstruction of both the quantum state itself as well as derived
quantities from it, e.~g.~purity, can be exactly determined.  This will
be illustrated in the next section.}

\subsection{Reconstruction of the Quantum State}

\begin{figure}
  \begin{center}
    \includegraphics[width = \columnwidth]{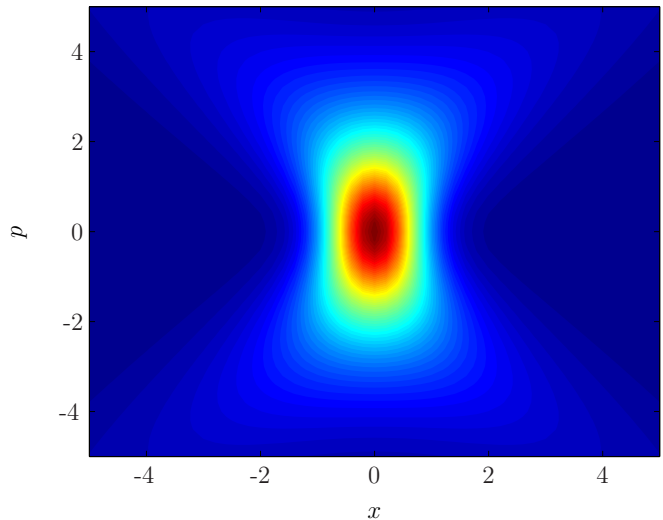}
    \caption{(Color Online) Reconstructed Wigner function: The Wigner
      function of the measured state as reconstructed using
Eq.~(\ref{sqz_phdiffused}) and the Markov chain.  This reconstruction contains
all the relevant statistical information.}
    \label{wignerfunction}
  \end{center}
\end{figure}

Having completely characterized the parameter space, the final state can be
reconstructed.  In order to generate the Wigner function shown in 
Fig.~\ref{wignerfunction}, the Markov chain together with
Eq.~(\ref{sqz_phdiffused}) were used to calculate the average Wigner function.
The characteristic non-Gaussian statistics of the 
phase-space Wigner distribution is clearly manifested in the reconstructed
state.  It should be stressed that this averaging already takes into account the
standard deviation of each value of the Wigner function since the parameter
values are taken from the posterior distribution.

\begin{figure}
  \begin{center}
    \includegraphics[width = \columnwidth]{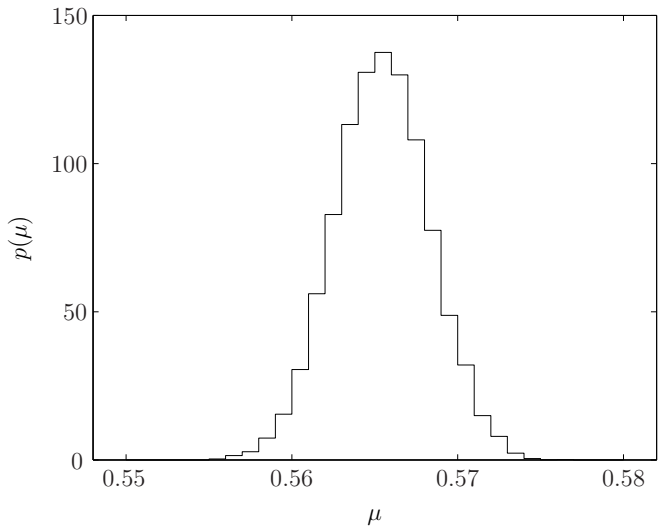}
    \caption{Reconstructed purity posterior distribution: The
      purity was calculated using the resulting Markov chain of 38,000 parameter
      values and Eq.~\parenth{\ref{purityformula}}.  This results in a probability
      density function which exactly quantifies the uncertainty on the estimation of
      the purity. The purity is calculated to be $\mu = 0.5649 \pm 0.0028$.  No such 
      result is possible with a pure maximum-likelihood approach.}
    \label{purity}
  \end{center}
\end{figure}

In addition to generating a phase-space plot of the reconstructed quantum state,
the Markov chain can be used in further calculations of such properties as the
purity of the reconstructed state.  Figure \ref{purity} depicts such a
result.  Using the analytical definition of the purity
\begin{equation}\label{purityformula}
\mu = 4 \pi \int \!\!\! \int \!\! W^2\parenth{x\,,p}\,dxdp, 
\end{equation}  
and the resulting chain from the MCMC, the purity can be calculated,
automatically taking into consideration the statistical error and correlation on
the parameters determined by the MCMC. The result is a probability density whose
standard deviation quantifies the degree of uncertainty on the estimation of the purity.
For the state in question, we obtain a purity of $\mu = 0.5649 \pm 0.0028$.
It is important to note that this information is delivered directly from the
MCMC itself; no additional assumptions as to the distribution of the errors and
their correlation properties need to be made. Additionally, any one-dimensional
quantity can be calculated in this manner.  For example, if estimating the
amount of entanglement of a non-Gaussian state, the logarithmic negativity
\cite{vidal2001} can be calculated over the span of the resulting chain.  The
result will be a probability density quantifying the uncertainty on its value.

\section{Conclusion}
We have applied a Bayesian data analysis scheme known as Markov chain Monte
Carlo (MCMC) to the tomographic reconstruction of quantum states.  Taking phase
diffused squeezed states as an example, we have provided the details as to the
derivation of the likelihood function as well as to the numerical
implementation of the MCMC. The results include a set of probability density
distributions which exactly quantify the degree of uncertainty on the estimation
of the parameters.  These results were compared to both a pure maximum-likelihood 
and a Fisher information approach.  Furthermore, using the Markov chain in the calculation of
the state's purity enabled the construction of a probability density
distribution on the value of the purity, thereby quantifying the degree of
uncertainty on its calculation.

We note that MCMC scheme is completely general and can be applied to higher
dimensional problems, such as the reconstruction of the density matrix, and will
be the topic of future publications. 

\begin{acknowledgments}
We would like to thank Christian R\"over for useful discussions and 
we acknowledge financial support from the Deutsche
Forschungsgemeinschaft (DFG), project number SCHN 757/2-1.  J.~F.~
acknowledges financial support from the Ministry of Education
of the Czech Republic under the projects Centre for Modern Optics
(LC06007) and Measurement and Information in Optics (MSM6198959213)
and from the EU under the FET-open grant agreement COMPAS (212008). 
\end{acknowledgments}

\begin{figure}[!t!]
\begin{center}
\includegraphics[width=\columnwidth]{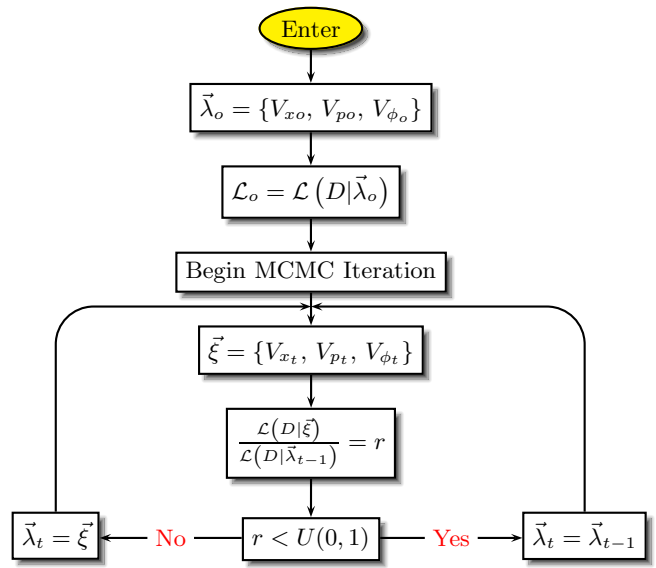}
\caption{(Color online) MCMC program flow chart: A schematic explanation of the
process of selecting MCMC samples from the posterior distribution.  The lower loop iterates
until the chain has converged to its equilibrium position and begins to
sample from this posterior distribution.}
\label{mcmc_progflow}
\end{center}
\end{figure}

\appendix

\section{Metroplis-Hastings Sampler}\label{mhsampler}
Our implementation of the MCMC is based on the Metroplis-Hastings sampler.  
It is defined by performing the following steps (see also
Fig.~\ref{mcmc_progflow}):

\begin{enumerate}
\item Generate initial values for parameters, $\vec{\lambda}_o$
\item Compute the quantity $\mathcal{L}_0 = \mathcal{L}(D|\vec{\lambda}_0)$
\item Iterate the following over the index $t$ until the chain has converged 

\begin{enumerate}
\item Generate a trial parameter vector $\vec{\xi}$ according to a proposal distribution $q(\vec{\lambda}_{t-1})$
\item Compute the quantity $\mathcal{L}_t = \mathcal{L}(D|\vec{\xi})$. If $V_xV_p<1$ or some $V_{j}<0$ then set $\mathcal{L}_{t}=0$.
\item Compute the ratio $r = \mathcal{L}_t/\mathcal{L}_{t-1}$
\item Sample from a uniform distribution, $U(0\,,1)$ \\
  if 
  $\left\{\begin{array}{cclr}
    r> U, &\mathrm{set}& \vec{\lambda}_{t} = \vec{\xi},\,\mathcal{L}_{t}=\mathcal{L}\,&\mathrm{accepted} \\ 
    r< U, &\mathrm{set}& \vec{\lambda}_{t}=\vec{\lambda}_{t-1},\,\mathcal{L}_{t}=\mathcal{L}_{t-1}\,&\mathrm{rejected}
  \end{array}\right.$

\end{enumerate}
\end{enumerate}

The proposal distribution $q(\vec{\lambda})$ used in stage 3(a) is used to select trial parameter 
values within the MCMC. Theoretically $q$ can be any distribution, however in practice it is 
sensible for the proposal distribution to suggest jumps that are local to the current location
but large enough to allow an efficient exploration of the parameter space~\footnote{There are many
techniques specific to the  efficient exploration of parameter space using MCMCs such as simulated annealing,
parallel tempering, and the Fisher information.  For the purposes of this analysis none of these
techniques were necessary and the proposal distribution used was an uncorrelated multi-variate
Gaussian with hand-tuned variances.}.  
Stage 3(b) ensures that the sampling is restricted to subspace of physically 
admissible values of parameters $\vec{\lambda}$.   
The simple difference between this MCMC method compared to that of a pure 
maximum-likelihood method is that for repeated stages within the Metropolis-Hastings
sampler the chain has finite probability of jumping \emph{both} to higher or lower 
values of likelihood.



\end{document}